\documentclass[11pt]{article}%
\usepackage{makeidx, siunitx, ragged2e}
\usepackage{booktabs,lscape,color, float}
\usepackage{subcaption} 
\usepackage{amsfonts,bm}
\usepackage{amsmath}
\usepackage{amssymb}
\usepackage{graphicx}
\usepackage{rotating}
\usepackage{multirow}%
\usepackage{geometry,setspace}
\usepackage{natbib}
\newtheorem{theorem}{Theorem}

\newtheorem{algorithm}[theorem]{Algorithm}

\newcommand{\R}{\mathbb{R}}
\newcommand{\E}{\mathbb{E}}
\newcommand{\V}{\mathbb{V}}

\begin{document}

\title{A Socioeconomic Well-Being Index}
\author{
A.~Alexandre Trindade\thanks{Texas Tech University, Department of Mathematics
\& Statistics, Lubbock TX 79409-1042, U.S.A., alex.trindade@ttu.edu.}
  \and
Abootaleb Shirvani\thanks{Texas Tech University, Department of Mathematics
\& Statistics, Lubbock TX 79409-1042, U.S.A., abootaleb.shirvani@ttu.edu (Corresponding
Author).}
  \and
Xiaohan Ma\thanks{Texas Tech University, Department of Economics, Lubbock TX 79409-1014, U.S.A., xiaohan.ma@ttu.edu.}
}
\maketitle

\begin{abstract}
An annual well-being index constructed from  thirteen socioeconomic factors is proposed
in order to dynamically measure the mood of the US
citizenry. Econometric models are fitted to the  log-returns of the
index in order to
quantify its tail risk and perform option pricing and risk budgeting. By
providing a statistically sound assessment
of socioeconomic content, the index is consistent with rational
  finance theory, enabling the construction and valuation of insurance-type financial
  instruments to serve as contracts written against it. Endogenously,
  the VXO volatility measure of the stock market appears to be
 the greatest contributor to tail risk. Exogenously,
``stress-testing'' the index against the politically important factors
of trade imbalance and legal immigration, quantify the systemic risk.
For probability levels in the range
of 5\% to 10\%, values of
trade below these thresholds are associated with larger downward
movements of the index than for immigration at the same level.
The main intent of the index is to provide  early-warning
for negative changes in the mood of citizens, thus alerting policy makers and private agents to
potential future market downturns.
\end{abstract}


\newpage
\section{Introduction}\label{sec:intro}
It is an obvious statement that financial market participants dislike
disruptions, especially those that are not based on economic
fundamentals. If an economy is likely to enter a recession, fully
informed and rational investors would store their wealth in more safe assets, or search for alternative investment opportunities in emerging
markets and elsewhere. Although the trough of a recession tends to materialize
	slowly enough to give investors time to prepare for it, opportunist or
uninformed traders, by contrast, may decide to ride the bubble until the
very end. For example, there were already signs of crash of 2008 almost one year ago. However, the overwhelming majority of market participants were content in simply riding the bubble in 2007 and early 2008. The tragedy of most investors then was that they decided to
upload their positions only 3 to 4 months ahead of the crash,
resulting in a "crowding effect" where all investors "rushed out of
the door", like in a bank run, thus creating a market avalanche. A consequence of this is that market crowding becomes an increasingly popular research topic. Although the reasons for it are now well known, few had any idea about its potential severity in 2007 before the Great Recession. 

Nowadays, potential sharp market downturns are studied very carefully. Regulators instituted
very strict standards with the Basel III Accord, and the capital requirements for firms that are
"too big to fail" are now quite severe. Even in high frequency
trading, flash-crashes are becoming rare. However, financial market
may still experience disruptions, due to, for example, an increase in
uncertainty associated with geopolitical events or polarization of
peoples opinion (e.g., the recent discussion concerning California’s
secession from the US), the negative impact of which cannot be
perfectly offset by policies. Moreover, policy makers themselves may
be as uninformed as private agents, and typically do not have perfect
information on the state of the economy. Policy decisions thus made
may not be effective in stabilizing the financial market, or in
serving as a useful signal for private agents.\footnote{For example, the current yield curve
	is ``bumpy'' suggesting that markets may be uncertain about future monetary policy.} Last but not
least, the mood (sentiment) of economic agents, as exemplified by the crowding
effect discussed earlier, may play a crucial rule in the stabilization of the financial market and the macroeconomy. 

The objective of this research is to employ state-of-the-art
statistically sound financial methods to construct a (to the best of
our knowledge) reliable and dynamic aggregate index based on a variety of
macro and micro economic factors, which will provide a quantitative
snapshot assessment of the US citizenry's level of socioeconomic
content.\footnote{We do not employ behavioral finance as it is not
  consistent with rational finance, and is thus unable to aid in the
  construction of insurance-type financial instruments to serve as
  (financial) contracts written against the index.} We term our
proposed index the Socioeconomic Well-Being Index (SWBI).\footnote{In
  contrast to many existing indices, the SWBI includes not only
  economic, but also diverse data related to social well-being.} While
also "stress-testing" against potential external factors like
immigration and trade imbalance, the index aims to determine the level
of future systemic risk, thereby serving as an "early-warning" mechanism for
serious potential undercurrent issues that could precipitate from changes
in the  mood of citizens, thus leading to potential future crises that
may be even more severe than in 2008. The SWBI assesses the tail risk (due to extreme events)
and provides forward-looking distributions for economic risk factors
and social well-being downturns. 

Specifically, the SWBI does this by being based on the
log-returns of  an equally-weighted linear combination of  factors.
Econometric ARMA-GARCH models driven by generalized hyperbolic noise,
satisfactorily capture the serial dependence and estimate the 
cross-sectional distribution of the data, as predicted by contemporary
best-practices financial theory \citep{massing2019best}. This framework also 
allows for the generation of Monte Carlo based future price scenarios,
leading to option pricing and risk budgeting for the SWBI. This enables the construction
and valuation of insurance-type financial instruments. Among the component series
 of the SWBI, the VXO volatility measure of the stock market is
 the greatest contributor to tail risk. 
Completing the rational finance-based valuation, stress-testing of the
SWBI against external factors like  trade imbalance and amount of legal
 immigration, quantifies the level of systemic risk. In this regard we
 find that the level of
 trade imbalance tends to be associated with a larger impact on
 negative well-being than does immigration.

There have been several recent noteworthy attempts by 
academics and non-academics alike  to quantify the well-being of the nation,
and/or to explore its implications on economic performance. Most
studies approach this issue via surveys, thereby
producing subjective indicators of contentment. For example, the
Gallup-Sharecare Well-Being Index is constructed from monthly telephone
interviews on how  people perceive and experience their
daily lives through five perspectives: purpose, social, financial,
community, and physical. The Gallup-Healthways Well-Being Index
interviews 1,000 US adults daily to provide real-time measurement
of health and well-being. The World Happiness Report combines various
global household surveys to construct three main happiness measures:
life evaluations, positive effect, and negative effect. The National
Accounts of Time Use and Well-being is used by \cite{Krueger_2009},
among others, to construct the subjective well-being of nations. 

The index we propose here differs from the above measures
by being based on time series of important macroeconomic aggregates,
which would arguably be more comprehensive, as well as immune from possible
response bias associated with subjective surveys\footnote{See \cite{McLean-2014} for a survey of national and international indices of
	well-being.}. A strand of the literature has already focused on such socioeconomic
factors affecting people's well-being. \cite{Blanchflower_Oswald_2004}
documents happiness trends in the US and Great Britain, and
quantitatively estimates the dollar values of events like unemployment
and divorce as sources influencing
happiness. \cite{Ferrer-i-Carbonell_Frijters_2004} develops a
conditional estimator for the fixed-effect ordered logit model to
re-evaluate the micro-level determinants of
happiness. \cite{DiTella_MacCulloch_Oswald_2003} shows that
macroeconomic movements, such as gross domestic product and
unemployment benefits, have significant impact on national
well-being. 

The main contribution of our paper, however, is in
constructing  a historical national well-being index based on
financial  econometric modeling and dynamic asset pricing theory. It
can therefore can be used for macroeconomic forecasting and the
issuing of marketable financial contracts, such as options and
futures.

  The rest of the paper is structured as follows. Section \ref{sec:data}
describes the
set of socioeconomic factors to be used to quantify the index, the
construction of which is detailed in Section \ref{sec:index}. This is
followed by econometric time series modeling of the index and 
its marginal density estimation in Section \ref{sec:econ}, this being a
prerequisite step for the option pricing and risk budgeting steps in Sections
\ref{sec:option} and \ref{sec:budgets}, respectively. The paper ends
with a stress-testing analysis in Section \ref{sec:stress}, where the
effect of external adverse socioeconomic factors on the tail risk is
examined. A discussion rounds out the paper. 


\section{Data Description}\label{sec:data}
The variables we select are those that have been shown in the literature to
affect the well-being of economic agents, such as in
\cite{Krueger_2009}, \cite{Blanchflower_Oswald_2004},
\cite{DiTella_MacCulloch_Oswald_2003},
\cite{Ferrer-i-Carbonell_Frijters_2004}, among others. Abbreviated
names for our list of 13 factors as well as their precise description
is as follows. (We append the prefix \texttt{Neg} to some of these if
we wish to also consider the reversed-sign version.)
\begin{description}
\item[\texttt{Confidence}.] The Consumer Confidence Index provides an indication of future developments of household consumption and savings, calculated based upon answers regarding their expected financial situation, their sentiment about the general economic situation, unemployment, and capability of savings.
\item[\texttt{CPI}.] Inflation as measured by consumer price index (CPI), is
  the growth rate of CPI for all urban consumers, which is a measure of the average
change rate in the price for goods and services paid by urban
consumers between any two time periods. (Reversed-sign version: \texttt{NegCPI}.) 
\item[\texttt{CrimeRate}.] Crime rate represents the estimated amounts of violent crimes per 100,000 people. (Reversed-sign version: \texttt{NegCrimeRate}.) 
\item[\texttt{DispIncome}.] Disposable income represents real disposable personal income, which is inflation adjusted personal income after payment of taxes. 
\item[\texttt{GDP}.] Real GDP (Gross Domestic Product) is the inflation adjusted value of
the final goods and services produced by labor and property located in
the United States. 
\item[\texttt{GenderParity}.] Gender parity is an index measuring the relative access to primary and secondary education for males and females, calculated as the quotient of the number of females to the number of males enrolled in the given stage of education. (Reversed-sign version: \texttt{NegGenderParity}.) 
\item[\texttt{GovTrans}.] Government transfer is the amount of government
social benefits provided to the unemployed, also known as unemployment
insurance. 
\item[\texttt{Inequality}.] Inequality is the Gini index of income inequality,
measuring household income dispersion.  (Reversed-sign version: \texttt{NegInequality}.) 
\item[\texttt{LifeExpect}.] Life expectancy indicates the number of
  years a newborn infant would live if prevailing patterns of
  mortality at the time of its birth were to stay the same throughout its life.  
\item[\texttt{Sentiment}.] The index of Consumer Sentiment is an economic indicator that measures how optimistic consumers perceive about their financial conditions and the state of the economy, constructed based on survey questions in the Survey of Consumers. 
\item[\texttt{Uncertainty}.] Uncertainty indicates the US policy uncertainty index based on
newspaper coverage frequency, the increase of which is found to
foreshadow declines in investment, output, and employment in the
United States.  (Reversed-sign version: \texttt{NegUncertainty}.) 
\item[\texttt{Unemploy}.] Unemployment represents the number of unemployed as a
percentage of the labor force (people 16 years of age and older, who
currently do not reside in institutions, and who are not on active duty
in the Armed Forces).  (Reversed-sign version: \texttt{NegUnemploy}.) 
\item[\texttt{VXO}.] The VXO index is the CBOE S\&P 500 Volatility
  Index, calculated by the Chicago Board Options Exchange
  (CBOE), and measuring the overall short-term volatility in the stock market. (Reversed-sign version: \texttt{NegVXO}.) 
\end{description}

Of these, real GDP, inflation, unemployment, government
transfer, life expectancy, disposable income, and VXO, are obtained
from the data set of the Federal Reserve Bank of St.~Louis. Inequality
is obtained from the US Census Bureau. Crime rate is calculated by the
FBI. The uncertainty index is developed by
\cite{Baker_Bloom_Davis_2016}. Consumer confidence is provided by OECD
(Organization for Economic Co-operation and Development) in its
publication of Main Economic Indicators: Business tendency and
consumer opinion surveys. Gender parity index is obtained from UNESCO
(The United Nations Educational, Scientific and Cultural
Organization). 

\begin{figure}[htb!]
  \begin{subfigure}[b]{0.5\textwidth}
    \includegraphics[width=\textwidth]{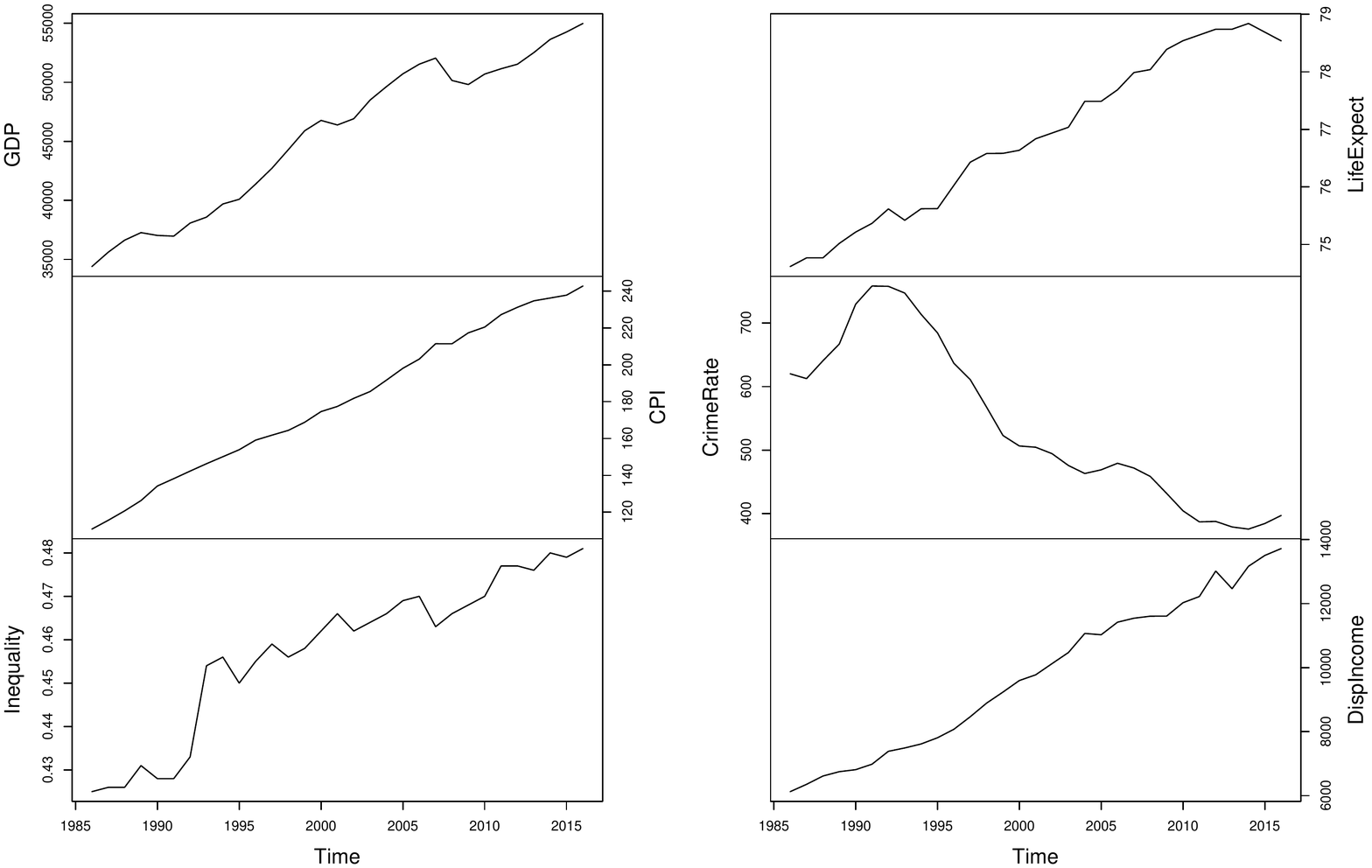}
    \caption{Trending.}
    \label{fig:FactorSeriesTrending}
  \end{subfigure}
  \begin{subfigure}[b]{0.5\textwidth}
    \includegraphics[width=\textwidth]{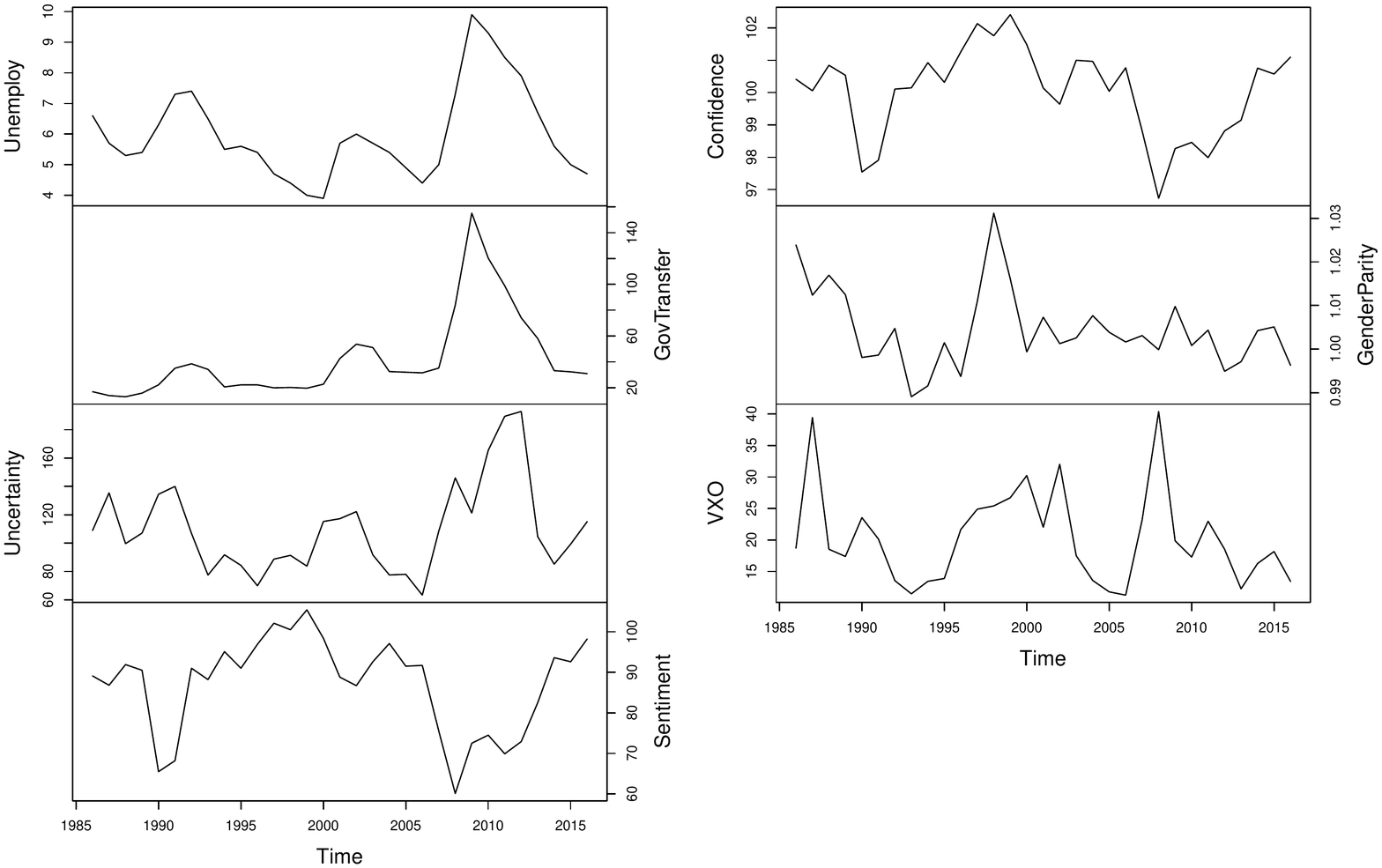}
    \caption{Stationary.}
    \label{fig:FactorSeriesStationary}
  \end{subfigure}
\caption{Time series of actual values for the factors with trending
  and with stationary behavior.}
\end{figure}

The common period of these 13 yearly factors is 1986-2016. 
Figures~\ref{fig:FactorSeriesTrending} and \ref{fig:FactorSeriesStationary} display
time series of their actual values. They seem to be of
two distinct types: \verb+GDP+, \verb+CPI+,
\verb+Inequality+, \verb+LifeExpect+, \verb+CrimeRate+,
and \verb+DispIncome+ display trending behavior, while the remaining 7
appear to be stationary. (However, at the 5\% level of significance the
Augmented Dickey-Fuller test does not reject the null hypothesis of
unit-root nonstationarity for any of the series.)

\begin{figure}[htb!]
\centering
\includegraphics[scale=0.5]{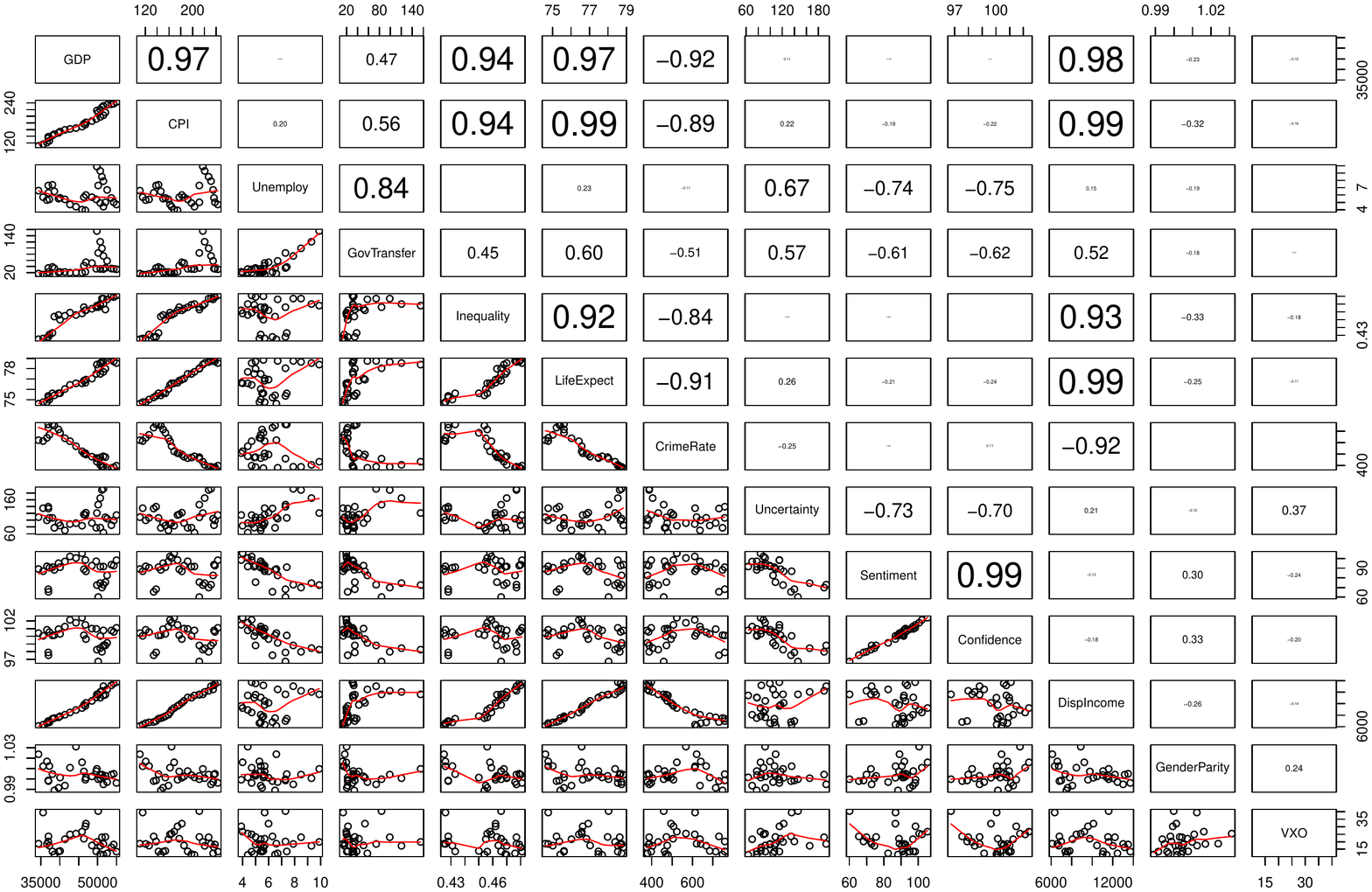}
\caption{Cross-sectional scatterplots of actual values for the 13 factors.}
\label{fig:Scatterplots}
\end{figure}

At this point we examined the existence of cointegration relationships
among the series in each of the two groups. For the trending group, the strong trends
coupled with small sample size does not allow for estimation of an underlying VAR
model necessary for carrying out Johansen's cointegration  test \citep{johansen1988statistical}. For
the stationary group, we find 2 cointegrating relationships
at the 5\% level of significance based on a VAR(2). (Higher order VAR
models could not be fitted due to the similar issue of collinearity and
small sample size.) Figure~\ref{fig:Scatterplots} displays cross-sectional
scatterplots of actual values for all 13 factors. We note that the large degree of
collinearity is easily spotted from the correlations on the
upper triangular portion, which are displayed with font size proportional to
the magnitude. There is therefore the question of whether all series
are necessary for the construction of the SWBI; an issue to be
explored in the next section.

\section{Construction of the Index}\label{sec:index}
In this section we detail the methodology for constructing the
SWBI. It employs the $13$ factors discussed in
the previous section. As is typical of macroeconomic variables, the Augmented Dickey-Fuller Test gives large p-values for all
series, confirming that they are all $I(1)$, or integrated of order
one. Thus, and in order to arrive at plausibly stationary series, we
transformed the actual values to log-returns, i.e., if $P(i,t)$ denotes the value of the
$i$-th factor at time $t$, then $r(i,t)=\log P(i,t)-\log P(i,t-1)$
denotes its log-return. 

\begin{figure}[htb!]
  \begin{subfigure}[b]{0.5\textwidth}
    \includegraphics[width=\textwidth]{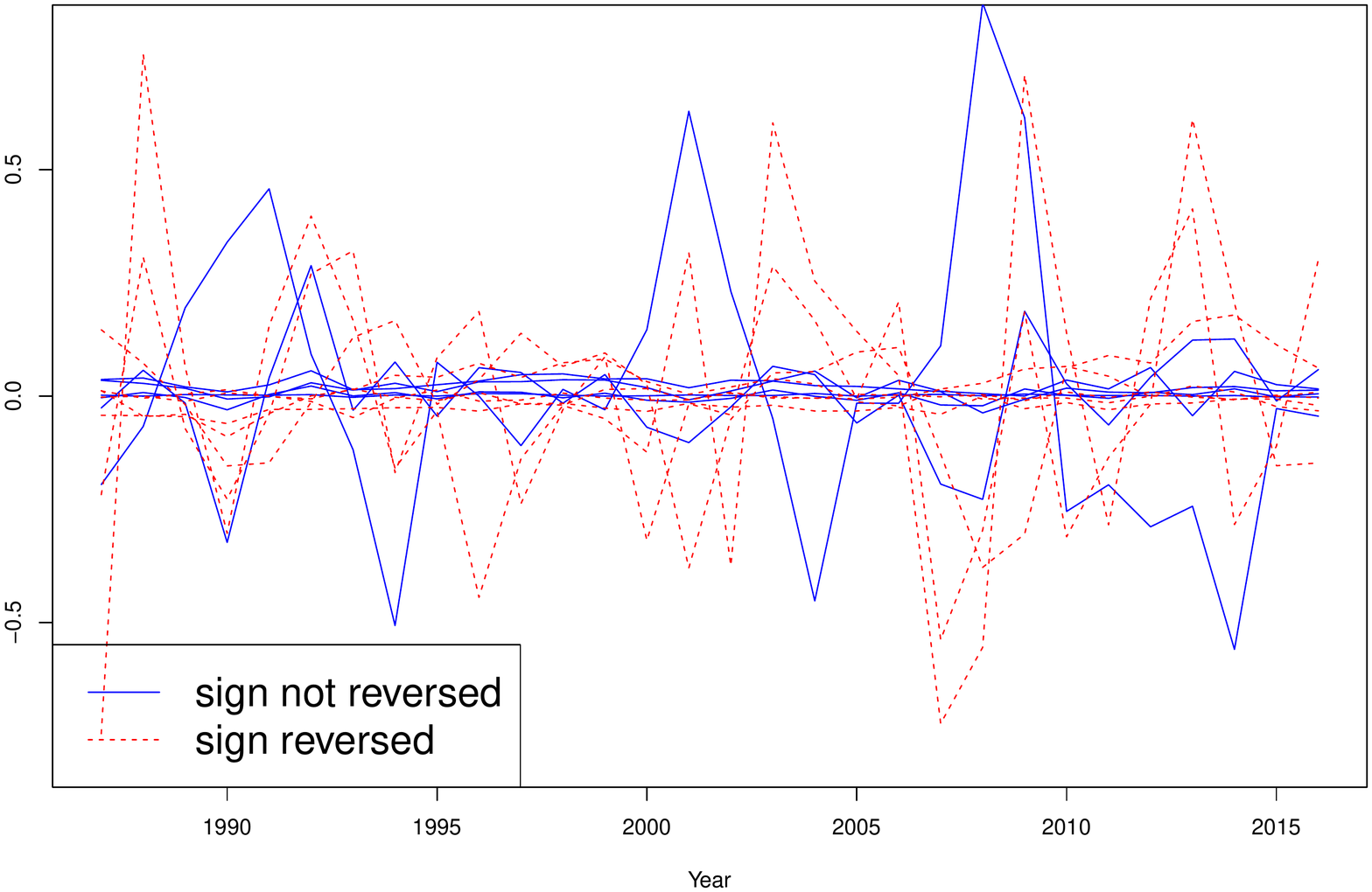}
    \caption{}
    \label{fig:FactorReturnSeries}
  \end{subfigure}
  \begin{subfigure}[b]{0.5\textwidth}
    \includegraphics[width=\textwidth]{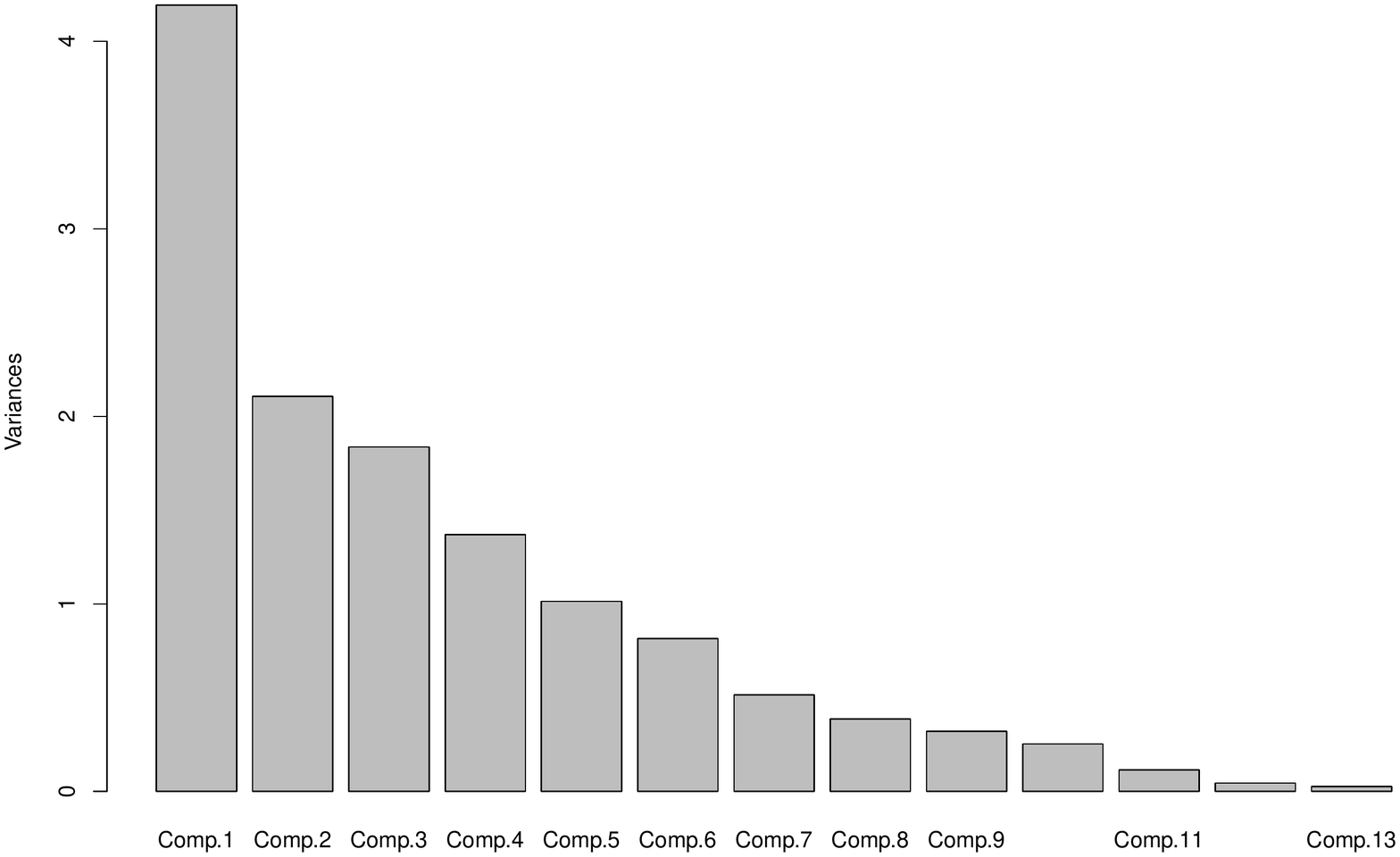}
    \caption{}
    \label{fig:PCAscreeplot}
  \end{subfigure}
\caption{Analysis for the panel of log-returns, $r(i,t)$, 
  defined in \eqref{eq:panel-rit}. (a) Time series plots with color and line type coded according to whether the sign was
  reversed (dotted) or not (solid). (b) Screeplot from a PCA giving the proportion of
  contribution from each component toward explaning the 13-dimensional
  correlation matrix.}
\end{figure}

In order to have positive values equate to greater well-being,  we
reversed the sign of the log-return value associated with the following factors: \verb+CPI+,
\verb+Unemploy+, \verb+Inequality+, \verb+CrimeRate+,
\verb+Uncertainty+, \verb+GenderParity+, and \verb+VXO+. On their
original scale, it can be argued
that large values of these factors would tend to be associated with
decreased well-being. We further set
the 1986 return value to zero for all factors, so that our starting
point is the panel of log-returns: 
\begin{equation}\label{eq:panel-rit}
\{r(i,t)\},\qquad i=1,\ldots,N=13, \quad t=1,\ldots,T=30.
\end{equation}
Time series plots of $r(i,t)$ for each of $i$ (each factor) are
displayed in Figure \ref{fig:FactorReturnSeries}. For comparison, the group whose sign
was reversed is shown in a different color and line type. All series
appear to be stationary now; at least with respect to trends and cycles.  

The formation of the SWBI value at time $t$, $r_t$, is
now obtained as an equally weighted linear combination of the $r(i,t)$ values for
each of the 13 factors at time $t$. More specifically, we implement
the following algorithm. 
\begin{algorithm}[Formation of SWBI]\label{algo:form-rt}
Starting from the panel of time series $r(i,t)$ in
\eqref{eq:panel-rit}, proceed as follows:
\begin{itemize}
\item[(i)] For each $i=1,\ldots,N$, standardize $r(i,t)$ according to its factor-level mean
  and standard  deviation: 
\[ 
R(i,t)=\frac{r(i,t)-m(i)}{s(i)}, \qquad
m(i)=\frac{1}{T}\sum_{t=1}^Tr(i,t), \quad s(i)=\frac{1}{T-1}\sum_{t=1}^T[r(i,t)-m(i)]^2.
\] 
\item[(ii)] Form the standardized index return series by weighting
  equally across all factors: 
\[
R(t)=\frac{1}{\sqrt{N}}\sum_{i=1}^NR(i,t). 
\]
\item[(iii)] Form the annual index return series by undoing the
  standardization in (i) according to the average of the factor means
  and standard  deviations: 
\[
r_t=m+s\;R(t), \qquad m=\frac{1}{N}\sum_{i=1}^Nm(i), \quad s=\frac{1}{N}\sum_{i=1}^Ns(i). 
\]
\end{itemize}
\end{algorithm}
For ease of reference in subsequent analyses, we call the resulting
$r_t$ series simply as the \texttt{Index} or SWBI. A time series plot is
displayed in the top left panel of
Figure~\ref{fig:rt-innovs-simulated}. 

At this point, and with
plausibly stationary factor log-returns as gleaned from Figure
\ref{fig:FactorReturnSeries}, one can properly investigate the issue
of whether all
13 factors that make up $r_t$ are needed. A principal components
analysis (PCA) on these  series suggests we need almost
all factors (correlation matrix based). Although the 1st PCA explains 32\% of the variability,
there is a very gradual contribution from each additional component,  
so that it is not until the 9th PCA that we obtain an explanatory
capability from these factors exceeding 95\%. Moreover, the weights in each PCA
are spread over almost all factors in each of these first 9 PCAs. In
summary, there is not an overwhelmingly clear indication that we should
reduce the dimensionality of the vector of 13 factors.  

\section{Econometric Analysis and Forecasting}\label{sec:econ}
\begin{figure}[htb!]
\centering
\includegraphics[scale=0.55, angle=0]{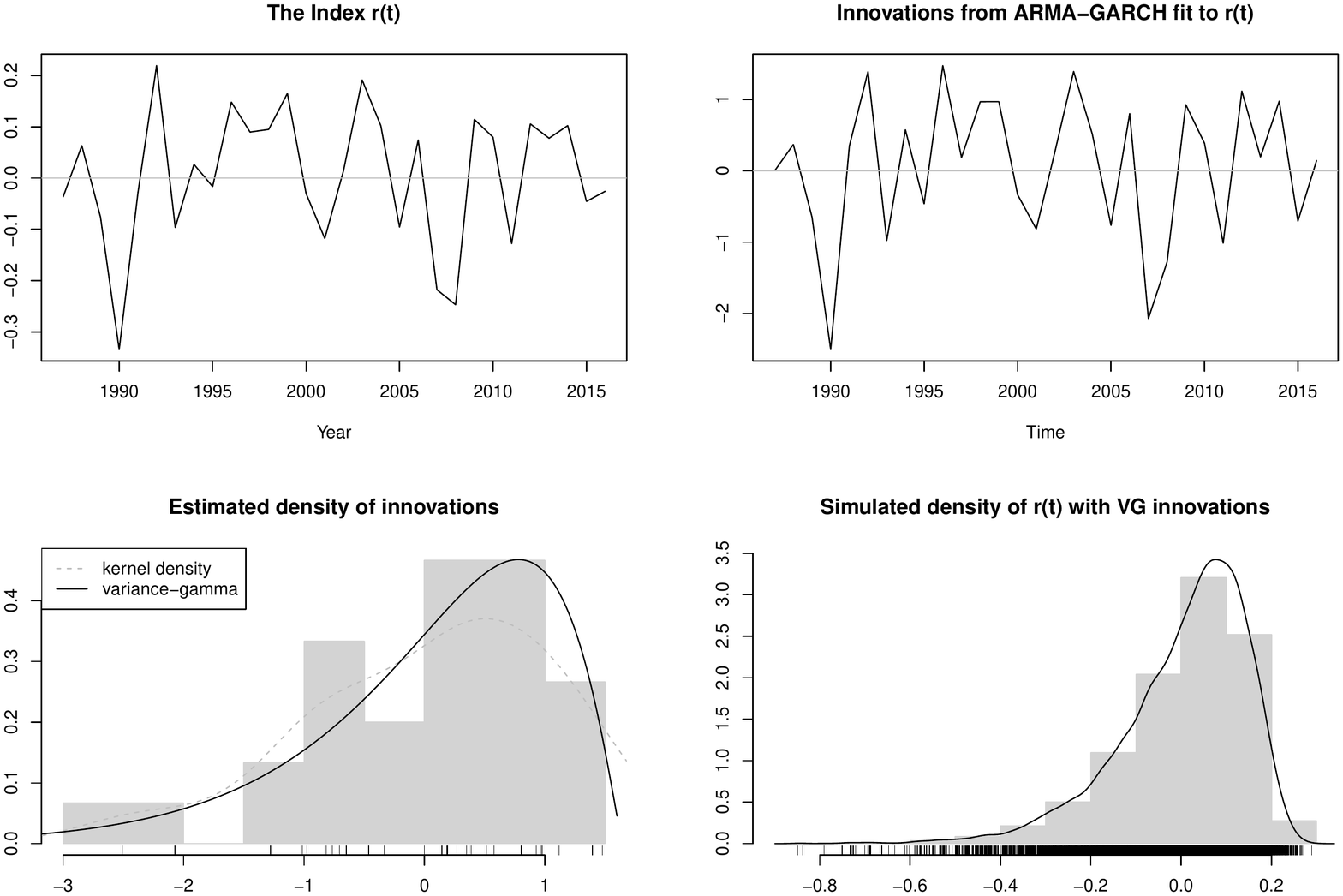}
\caption{Time series of the \texttt{Index} $r_t$ and its innovations from an
  ARMA-GARCH fit (top panels). The bottom right panel displays 10,000
  scenarios from the marginal of $r_t$ based on a
  VG distribution fitted to the ARMA-GARCH innovations (bottom left panel).}
\label{fig:rt-innovs-simulated}
\end{figure}

In this section we fit time series models to $r_t$. This will allow
for evaluation of risk measures and the performing of option pricing  for the
\texttt{Index}. Two models were entertained here. Searching for the best-fitting ARIMA model via AIC and BIC
points to an ARIMA(0,0,0) with zero mean, i.e., white noise. However, since the option pricing
relies on an ARMA-GARCH configuration, we fit also an
ARMA(1,1)-GARCH(1,1) with zero mean and normal innovations. The
innovations from this fit (the raw
residuals divided by the GARCH conditional standard deviation estimate) are displayed in the top right panel of
Figure~\ref{fig:rt-innovs-simulated}. 

The generation of multiple scenarios from the (stationary distribution
of the) fitted model for $r_t$ in order to calculate
risk measures, a step we loosely call ``forecasting'', can be
performed once an appropriate model for the innovations has been determined. 
The standard parametric family in this context is
the \emph{Generalized Hyperbolic} (GH) distribution, originally
  introduced by \cite{BN77}. \cite{jorgensen1982statistical} is the classical reference
  on its properties, while an accessible overview is given
  by \cite{paolella2007intermediate}. Briefly, the GH features both
  heavy tails and skewness, includes the familiar elliptical family as
  a special case (normal, t, etc.), exhibits tail-dependence
  \citep{rank2007copulas}, and is infinitely divisible, a necessary
  and sufficient condition to build Levy processes (ubiquitous in
  financial time-series due to their continuity and ability to model
  jumps). For these and other reasons it has become the default
  distribution in  modeling the returns of equity indices at various
  temporal scales \citep{massing2019best}.

Notationally, we designate by $X\sim\mathcal{GH}\left(\lambda,\alpha,\beta,\delta,\mu\right)$
a random variable following a GH with parameters $\lambda\in\R$ (tail heaviness), $\alpha>0$
(shape), $\beta\in\R$ (skewness) such that $\alpha^2-\beta^2>0$, $\delta>0$ (scale), and $\mu\in\R$ (location). 
(There are at least two alternative parametrizations in common usage). The density function is derived by
mixing normals according to a Generalized Inverse Gaussian
distribution (a special case of GH), and does not therefore in general
have a closed-form representation. Software
implementation of the GH is provided through the R
library \texttt{ghyp} \citep{Rlib-ghyp} and its accompanying vignette.
Two special cases of the GH we focus on in this study are the
Variance Gamma (VG), which is a GH with $\lambda>0$ and $\delta=0$, and the Negative Inverse
Gaussian (NIG), a GH with $\lambda=-1/2$; both of these possessing exponential tails. The latter in particular is
pointed out by \cite{BN97} and \cite{BN2007nig} as being especially
appropriate for stochastic volatility modeling based on a
homogeneous Levy process, and the option pricing models
that can consequently be constructed.

Fitting VG and NIG marginal models to the sample of
ARMA(1,1)-GARCH(1,1) innovations seen as the rug and histogram plot in the lower left panel of
Figure~\ref{fig:rt-innovs-simulated}, leads to the VG density displayed as the
solid line in
that figure (the NIG being an inferior fit). The appropriateness of the fit is
confirmed by the Kolmogorov-Smirnov and Anderson-Darling tests with p-values exceeding
0.9. Simulating 10,000 scenarios from the overall fitted ARMA(1,1)-GARCH(1,1)
model with VG innovations, then leads to the marginal of $r_t$ seen
as the rug and histogram in the lower right panel of
Figure~\ref{fig:rt-innovs-simulated}; the solid line being a kernel
density smoother. Summary statistics and left tail risk measures for
these scenarios are shown on Table~\ref{tab:summary-risk}. Here, the
Value-at-Risk (VaR) and expected shortfall (ES) risk measures are reported \citep[Ch.~2]{ch2007modeling}.

\begin{table}[!htb]
\centering
\caption{Summary statistics and left tail risk measures for the 10,000
  scenarios generated from the ARMA(1,1)-GARCH(1,1)
model with VG innovations fitted to the
\texttt{Index}.}
\label{tab:summary-risk}
\begin{tabular}{ccc}
\toprule
 \multicolumn{3}{c}{Summary Statistics} \\
  Minimum/Maximum & Mean/Median & Skewness/Kurtosis (excess)\\
  \midrule
 -0.8492/0.2894 & 0.0064/0.0351 &  -1.181/2.094 \\
  \bottomrule
 \multicolumn{3}{c}{Left Tail Risk Measures} \\
1\% VaR/ES & 5\% VaR/ES & 10\% VaR/ES \\
  \midrule
  -0.4411/-0.5458 & -0.2655/-0.3720 & -0.1826/-0.2960 \\
\end{tabular}
\end{table}

\section{Option Pricing}\label{sec:option}
Options can be used for hedging, speculating, and calculating the risk
of investments. 
The most common option pricing models are Black-Scholes, Binomial,
Trinomial tree, Monte-Carlo simulation, and finite
difference. Recently, the discrete stochastic volatility based model
has received considerable attention, particularly with regard to explaining some well-known mispricing phenomena. 
The pricing model in a discrete-time conditional heteroskedasticity setting was first
considered by \cite{Duan:1995}. He used a GARCH driven by normal
innovations for asset returns in order to price options. We follow his
strategy by considering a standard GARCH model with GH innovations to
calculate the fair value of an option of our \texttt{Index}
(SWBI). Specifically, we assume the log-returns $r_t$ follow the process:
\begin{equation}
\label{Dynamic_procss_WBI}
r_t=\log{\frac{I_t}{I_t-1}}=r'_t+\lambda_0 \sqrt{h_t}-\frac{1}{2}h_t+\sqrt{h_t} \epsilon_t,
\end{equation}
where $h_t=\V\left( r_t \mid F_{t-1}\right)$ is the conditional
variance at time $t$, with $F_{t-1}$ denoting the information
set consisting of all linear functions of past returns available
up to time $t-1$, $r'_t$ is the risk-less rate of return at time $t$, and $\lambda_0$ is the risk premium for the SWBI.  We use a GARCH(1,1) with GH innovations to model the conditional variance as follows:
\begin{equation}\label{GARCH_model}
h_t=m+a\, h_{t-1}+b\, \epsilon_{t-1}^{2},\qquad
\{\epsilon_t\}\sim\text{iid }\mathcal{GH}\left(\lambda,\alpha,\beta,\delta,\mu\right). 
\end{equation}
\cite{Blaesild:1981} proved that under this model, the conditional distribution of $r_t$
given $F_{t-1}$ on the \emph{real-world} probability space $\mathbb{P}$ is distributed as
\begin{equation}\label{WBI_distribution}
r_t \sim
\mathcal{GH}\left(\lambda,\frac{\alpha}{\sqrt{h_t}},\frac{\beta}{\sqrt{h_t}},\delta\sqrt{h_t},r'_t+m_t+\mu
  \sqrt{h_t} \right) ,\qquad\text{with } m_t=\lambda_0 \sqrt{h_t}-\frac{1}{2}h_t.
\end{equation}

Options are priced on the risk-neutral probability (probability space
of future outcomes adjusted for risk). In an incomplete market to
price options, the crucial issue is to identify an equivalent
martingale measure to obtain a consistent price for contingent
claim. Using the \cite{Gerber:1994} Esscher transformation,
\cite{Chorro:2012} proved that, under model \eqref{GARCH_model}, the conditional distribution of $r_t$ given $F_{t-1}$ on the \emph{risk-neutral} probability space $\mathbb{Q}$ is  distributed as
 \begin{equation}
\label{WBI_Q_distribution}
r_t \sim \mathcal{GH}\left(\lambda,\frac{\alpha}{\sqrt{h_t}},\frac{\beta}{\sqrt{h_t}}+\theta_t,\delta \sqrt{h_t},r'_t+m_t+\mu \sqrt{h_t} \right) ,
 \end{equation}
where, and with $M(\cdot)$ denoting the conditional
moment generating function of $r_{t+1}$ given $F_{t}$ on $\mathbb{P}$, $\theta_t$ solves the equation 
\begin{equation}\label{MGF_distribution}
M\left(1+\theta_t \right)=M\left(\theta_t \right)\exp\{r'_t\}.
\end{equation}
(See
\cite{Duffie:2001} for details on
the precise definitions of the probability spaces $\mathbb{P}$ and $\mathbb{Q}$.)

To price the SWBI call option up to a given \emph{time to maturity} $T$, we use Monte Carlo simulation to generate future values of SWBI via the scheme implemented by \cite{Chorro:2012}, as follows:
\begin{enumerate}
	\item Fit the GARCH(1,1) model
          \eqref{GARCH_model} to the available historical log-returns of SWBI.   
	\item Set $t=0$ and forecast $h_1$, the conditional variance at time $t=1$.
	\item Starting from $t=1$, repeat steps (a)--(c) below under $\mathbb{Q}$
          for $t=1,\ldots,T$:
	 \begin{enumerate}
		\item solve \eqref{MGF_distribution} to find $\theta_t$;
		\item generate $\epsilon_{t+1}$ from the stationary
                  distribution of $\epsilon_t\sim\mathcal{GH}(\lambda,\alpha,\beta+\sqrt{h_t}\theta_t,\delta,\mu);
$  
		\item compute $r_{t+1}$ and $h_{t+1}$.
	\end{enumerate}
	\item This scheme yields $\{r_1,\ldots,r_T\}$ under
          $\mathbb{Q}$. Thus the future price of the SWBI at time $T$
          is given by 
	\begin{equation}
	 	I_T=I_0\exp\left\{\sum_{t=1}^{T}r_t\right\}.
	 \end{equation}
	 \item Repeat step (3)--(4) in order to simulate $N=10,000$
           future price values of the SWBI: $\{I_T^{(1)},\ldots,I_T^{(N)}\}$. 
\end{enumerate}
The approximate call $(\hat{C})$ and put $(\hat{P})$ option prices at time $t\leq T$ are then
computed, for a given \emph{strike price} $K$, as the Monte Carlo averages:
	 \begin{eqnarray*}
	\hat{C}\left(t,T,K
           \right) &=& \,e^{-r'_t(T-t)}\frac{1}{N}\sum_{i=1}^{N}\left(I_T^{(i)}-K
           \right)_{+}, \\ 
	\hat{P}\left(t,T,K \right) &=& \,e^{-r'_t(T-t)}\frac{1}{N}\sum_{i=1}^{N}\left(K-I_T^{(i)} \right)_{+}.
	 \end{eqnarray*}

\begin{figure}[htb!]
	\centering
	\includegraphics[scale=0.35]{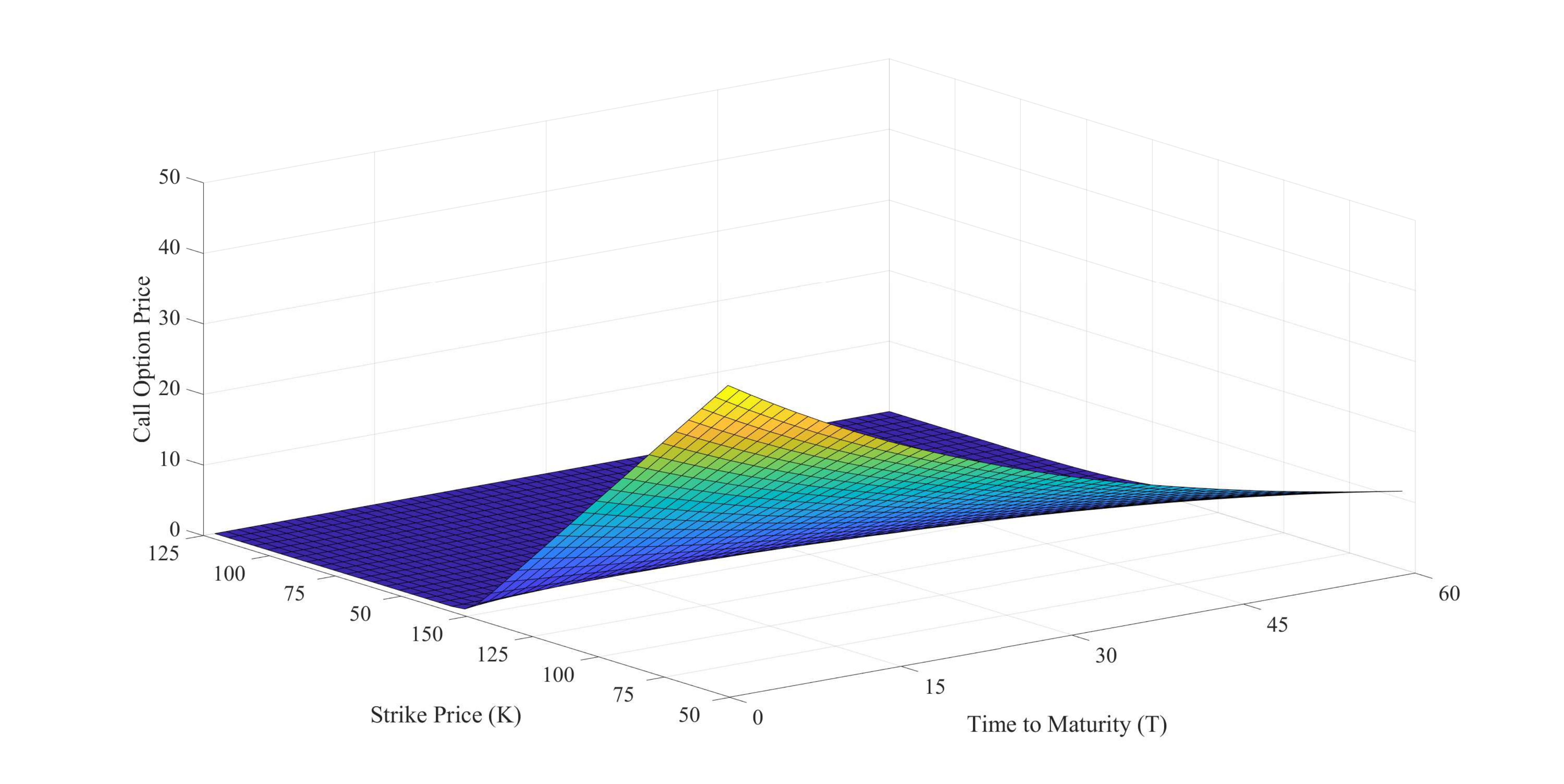}
	\caption{Call-Option prices against time to maturity and strike price.}
	\label{fig:Call_option_price}
\end{figure}

\begin{figure}[htb!]
	\centering
	\includegraphics[scale=0.35]{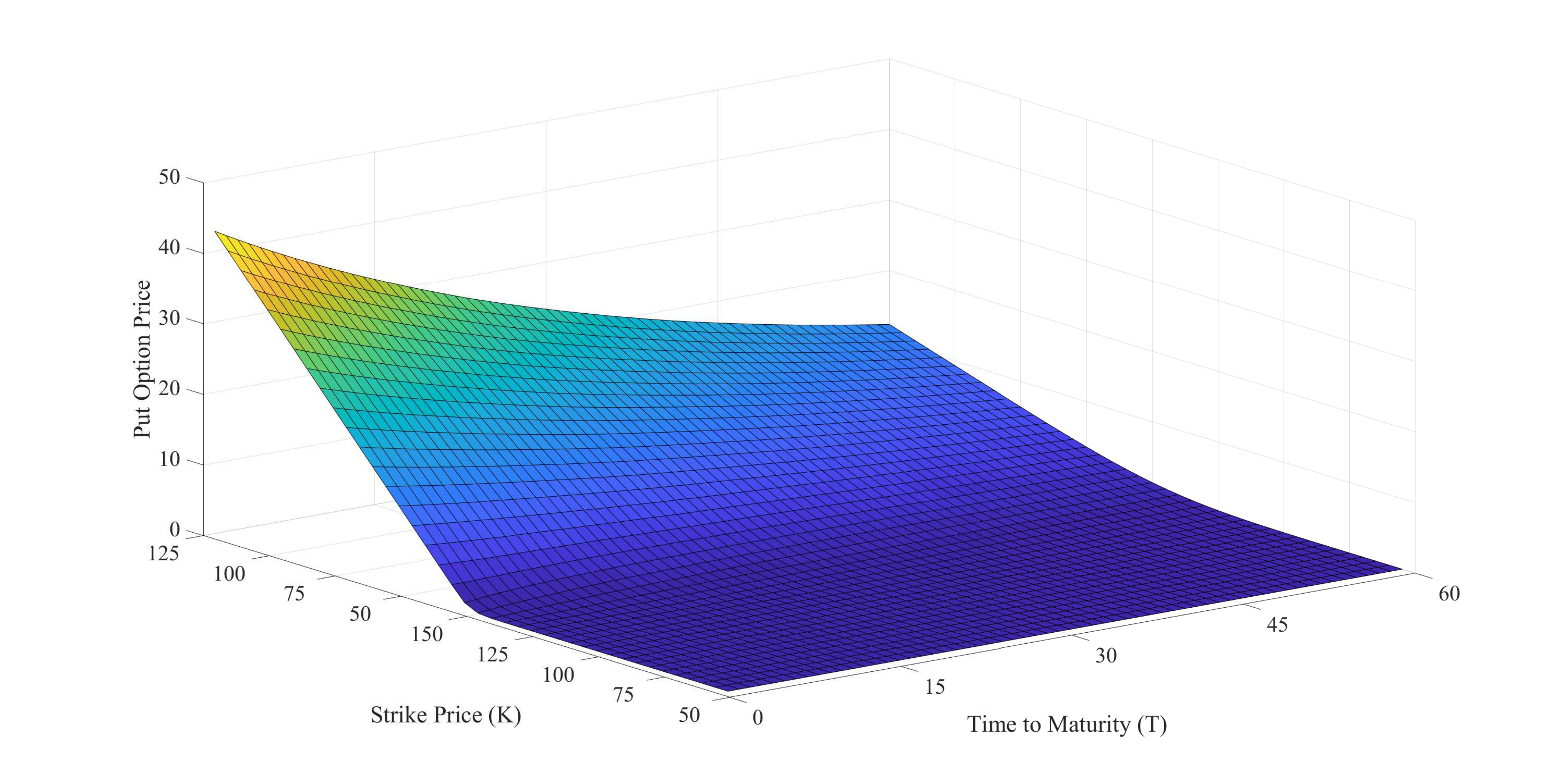}
	\caption{Put-Option prices against time to maturity and strike price.}
	\label{fig:Putt_option_price}
\end{figure}

\begin{figure}[htb!]
	\centering
	\includegraphics[scale=0.35]{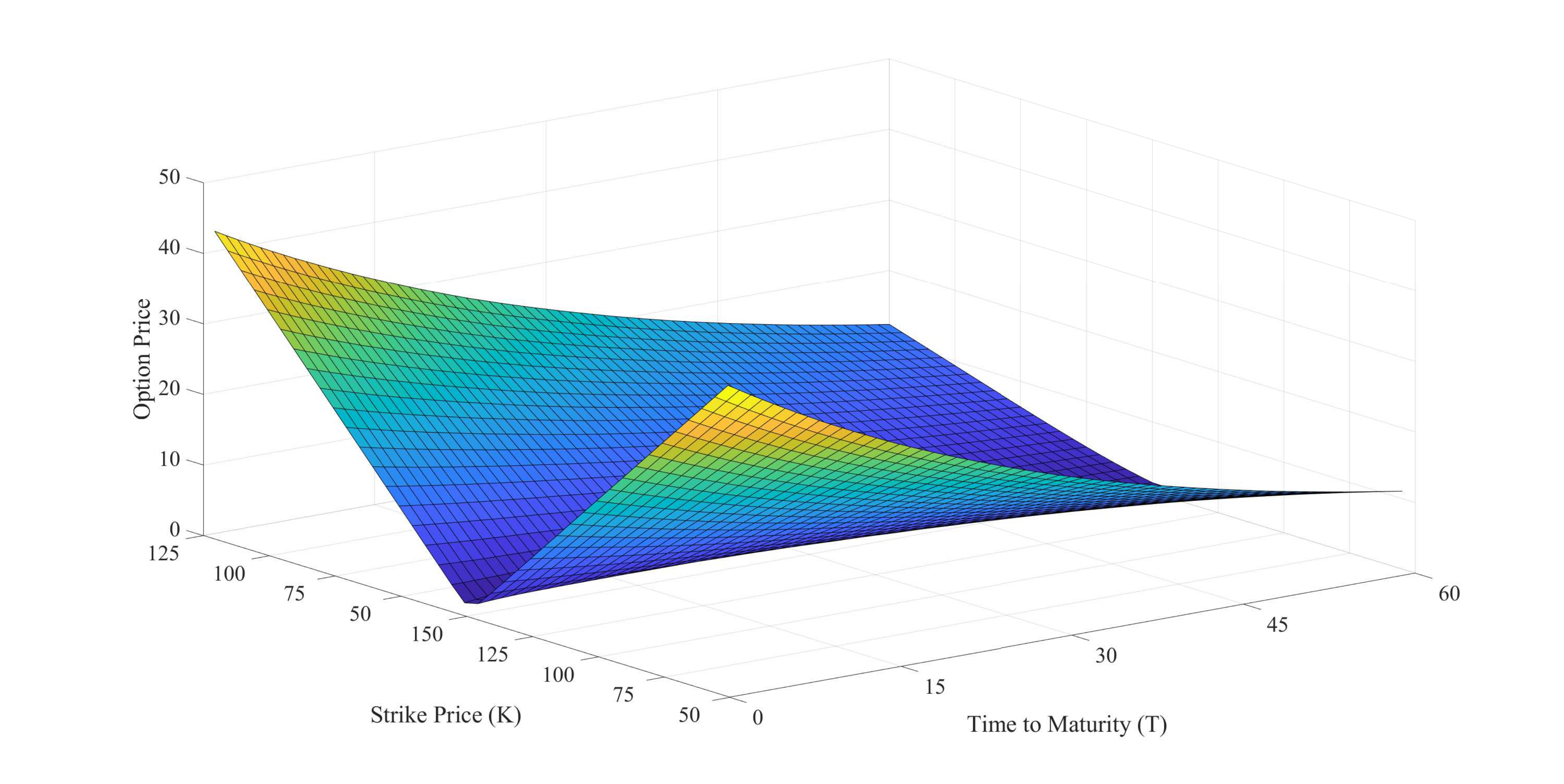}
	\caption{Call and Put-Option prices against time to maturity and strike price.}
	\label{fig:option_price}
\end{figure}

\begin{figure}[htb!]
	\centering
	\includegraphics[scale=0.35]{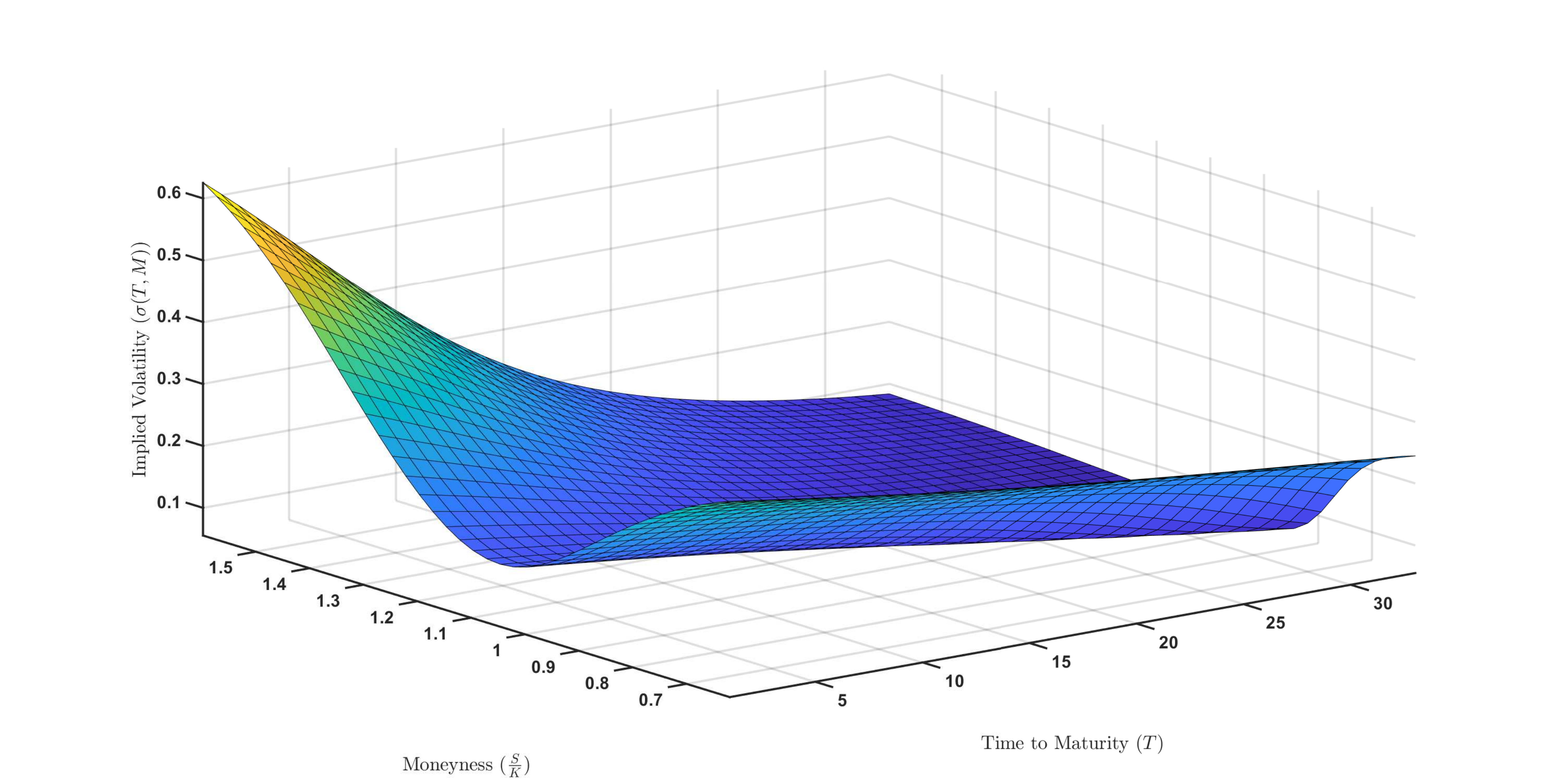}
	\caption{Implied volatility against time to maturity and Moneyness.}
	\label{fig:Implied_vol}
\end{figure}

Figures \ref{fig:Call_option_price}-\ref{fig:option_price} show call
option prices, put option prices, and call-put option prices,
respectively, plotted against both maturity ($T$) and strike
($K$). The graphs reveal the relationships among time to maturity,
strike, and option prices. As expected, for a fixed $K$ we see a
decline in price as $T$ increases. Figure
\ref{fig:Implied_vol} plots the implied volatility surface (the
market's view of the future value of volatility) against time to
maturity and \emph{Moneyness} (defined as $S/K$, where $S$ is the price of
the stock). Again as expected, we observe that as time to maturity decreases, the
volatility surface increases, reflecting the fact that a higher level
of uncertainty exists regarding whether or not the
option will be exercised. 


\section{Risk Budgets}\label{sec:budgets}	
Risk budgets are often used to allocate the risk of a portfolio by
decomposing the total portfolio risk into the risk contribution of
each component position. Portfolio standard deviation (Std),
Value-at-Risk (VaR), and expected tail loss (ETL) budgets are the most
popular strategies used to better understand the center-risk and
tail-risk contributions.  \cite{Chow:2001}, \cite{Litterman:1996},
\cite{Maillard:2010}, and \cite{Peterson:2008} studied the use of
portfolio Std and VaR in risk budgeting. \cite{Boudt:2013} reviewed
the ETL budgets. Here, we use the Std and ETL risk budgets as an investment
strategy under the condition of an equal-weights portfolio.  The
equal-weights portfolio is widespread in practice because it does not
require information on the risk and return, and supposedly
provides a diversified portfolio. 

Thus, we have the vector of portfolio weights
${w}=(w_1,\ldots,w_N)$ at each time point $t$, where $N=13$ and $t=1,\ldots,31$. We  first
define the marginal risk and risk contribution of the $i$th asset in
the portfolio, and then calculate the respective Std and ETL risk contributions.   
Let $R(w):\mathbb{R}^n \rightarrow \mathbb{R}$ denote a risk measure in the portfolio weight vector, ${w}$, then the marginal risk contribution of the $i$-th asset  denoted by $\textit{RC}_i$ is 
\begin{equation}
\label{MRCT}
\textit{RC}_i(w) = w_i \frac{\partial R(w)}{\partial w_i}.
\end{equation}
The marginal risk contribution of the $k$-th subset is 
\begin{equation}
\label{PCTR}
\textit{RC}_{M_k}(w) = \sum_{i \in M_k}^{}\textit{RC}_i (w),
\end{equation}
where $M_k\subseteq\left\lbrace 1,\ldots,N \right\rbrace$ for each $k=1,\ldots,s$, denotes $s$ subsets of portfolio assets. 

Table \ref{tab:risk_budget} reports the estimated risk allocation of the equal-weights portfolio. It seems \texttt{NegVXO} has a relatively higher risk than the other factors. Meanwhile, \texttt{GovTrans} factor has the lowest tail risk contribution in both cases, and \texttt{NegUnemploy} has the lowest center risk contributions. Thus, the tail risk diversifiers are \texttt{GovTrans}, \texttt{DispIncome}, \texttt{GDP}, \texttt{NegInequality}, \texttt{NegCrimeRate}, \texttt{NegGenderParity}, and \texttt{LifeExpect}. The tail risk contributors are the remaining factors. Note that \texttt{NegVXO} is the main risk contributor among all factors. 
\begin{table}[htb!]
	\caption{ Standard deviation and ETL Risk Budget}
	\centering
	\begin{tabular}{@{}lcccccc@{}}
		\toprule
		\multicolumn{1}{l}{Factors}
          & \begin{tabular}[c]{@{}c@{}}MCTR\\ ETL (95)\end{tabular}
          & \begin{tabular}[c]{@{}c@{}}PCTR \\ ETL (95)\end{tabular} & \begin{tabular}[c]{@{}c@{}}MCTR\\ ETL (99)\end{tabular} & \begin{tabular}[c]{@{}c@{}}PCTR\\ ETL(99)\end{tabular} & \begin{tabular}[c]{@{}c@{}}MCTR\\ (Std)\end{tabular} & \begin{tabular}[c]{@{}c@{}}PCTR\\ (Std)\end{tabular} \\ \midrule
		GovTrans                   & -0.70\%                                                 & -7.25\%                                                 & -1.17\%                                                 & -9.38\%                                                & 3.13\%                                               & 4.75\%                                               \\
		DispIncome                     & -0.10\%                                                 & -1.07\%                                                 & -0.03\%                                                 & -0.27\%                                                & -0.12\%                                              & -0.19\%                                              \\
		GDP                         & -0.08\%                                                 & -0.83\%                                                 & -0.07\%                                                 & -0.59\%                                                & 0.22\%                                               & 0.33\%                                               \\
		NegInequality                     & -0.06\%                                                 & -0.62\%                                                 & -0.11\%                                                 & -0.87\%                                                & -0.24\%                                              & -0.36\%                                              \\
		NegCrime                    & -0.03\%                                                 & -0.31\%                                                 & 0.04\%                                                  & 0.32\%                                                 & 0.09\%                                               & 0.13\%                                               \\
		NegGenderParity                   & -0.03\%                                                 & -0.26\%                                                 & -0.02\%                                                 & -0.19\%                                                & -0.24\%                                              & -0.37\%                                              \\
		LifeExpect                   & -0.02\%                                                 & -0.24\%                                                 & -0.03\%                                                 & -0.23\%                                                & -0.02\%                                              & -0.03\%                                              \\
		Confidence                    & 0.13\%                                                  & 1.36\%                                                  & 0.18\%                                                  & 1.44\%                                                 & 0.73\%                                               & 1.11\%                                               \\
		NegCPI                      & 0.32\%                                                  & 3.26\%                                                  & 0.38\%                                                  & 3.08\%                                                 & 0.19\%                                               & 0.29\%                                               \\
		NegUnemploy                    & 0.60\%                                                  & 6.14\%                                                  & 1.02\%                                                  & 8.18\%                                                 & -0.31\%                                              & -0.46\%                                              \\
		Sentiment                    & 1.33\%                                                  & 13.69\%                                                 & 1.78\%                                                  & 14.27\%                                                & 7.93\%                                               & 12.04\%                                              \\
		NegUncertainty                   & 3.19\%                                                  & 32.84\%                                                 & 4.01\%                                                  & 32.16\%                                                & 19.87\%                                              & 30.18\%                                              \\
		NegVXO                      & 5.17\%                                                  & 53.30\%                                                 & 6.50\%                                                  & 52.09\%                                                & 34.63\%                                              & 52.58\%                                              \\ \bottomrule
	\end{tabular}
	\label{tab:risk_budget}
\end{table}

\section{Stress-Testing}\label{sec:stress}
Asset and investment management firms commonly use \emph{stress-testing} to determine the resilience of a given portfolio against
possible undesirable financial situations (assess the risk), and then set in place any hedging
strategies necessary to mitigate against possible
losses. The intent is to evaluate how well the assets
might weather certain market occurrences and external events. The
determination of appropriate factors that may contribute to these
``stressful'' events, is in itself a difficult task.

In this section we consider stress testing the \texttt{Index} with the contemporaneously observed factors \emph{trade
  balance} (\texttt{Trade}) and \emph{legal
  immigration} (\texttt{Immig}), acting as stressors. The data on \texttt{Trade} represents the difference between the total value of
US export and import of goods and services, based on the Balance of
Payments. The annual data is obtained from the US Census
Bureau. \texttt{Immig} is the total number of persons obtaining lawful
permanent resident status, and is retrieved from the Department of
Homeland Security. We choose these two variables as stress factors
because they may be sensitive to policy makers' decisions, but they
should be \emph{a priori} neutral  in that changes in  their
values do not straightforwardly imply a cause-effect relationship with
the well-being of US citizens. 

\begin{figure}[htb!]
	\centering
	\includegraphics[scale=0.6]{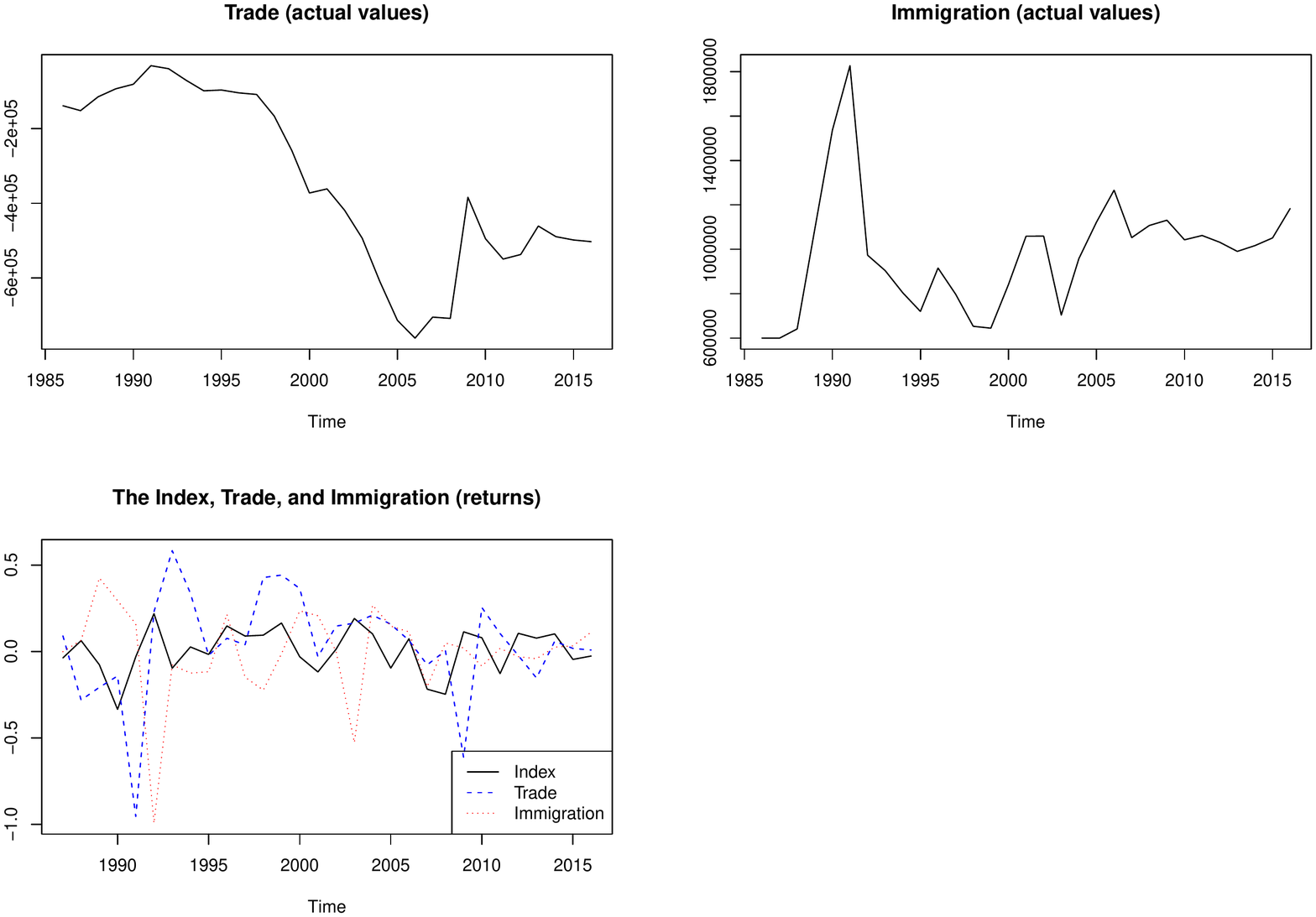}
	\caption{Time Series Plots of the \texttt{Index} and stress
          factors \texttt{Trade} and \texttt{Immig}.}
	\label{fig:Index-Trade-Immig}
\end{figure}

The top panels of
Figure~\ref{fig:Index-Trade-Immig} plots the actual values of the stress factor time series,
while the bottom panel shows the log-returns of these as well as the
\texttt{Index}, which are all white noise, as expected. In fact, a Ljung-Box test on all 3
series of returns does not detect any serial correlation, or
even dependence (all p-values larger than 0.5). Thus we opted not to put
the series through the ARMA(1,1)-GARCH(1,1) filter for this analysis,
i.e., the ARIMA(0,0,0) with zero mean suggested by AIC and BIC in
Section~\ref{sec:econ} was fitted to the returns of all three series:
\texttt{Index}, \texttt{Trade}, and \texttt{Immig}. 

\begin{figure}[htb!]
  \begin{subfigure}[b]{0.5\textwidth}
    \includegraphics[width=\textwidth]{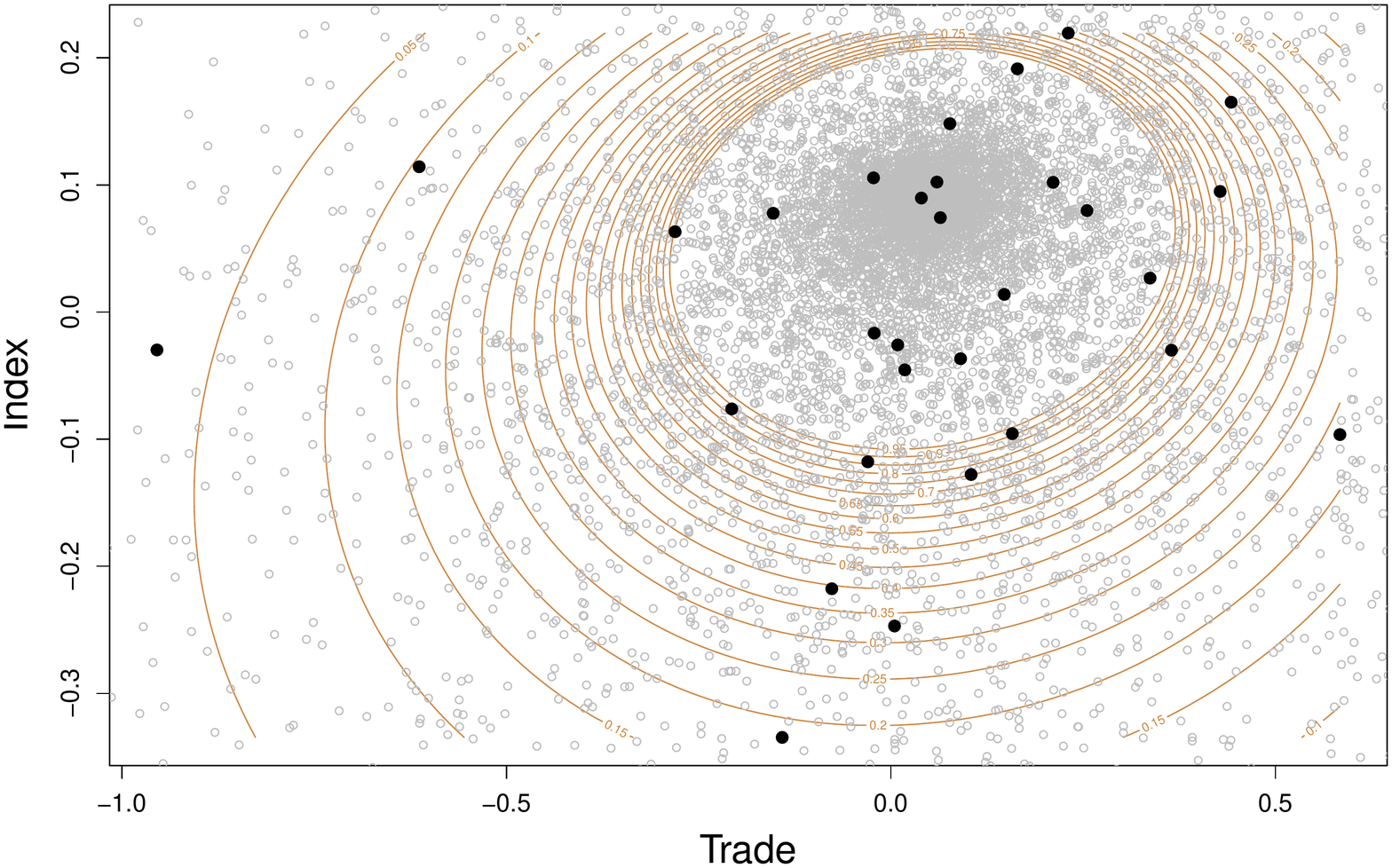}
    \caption{\texttt{Trade} vs.~\texttt{Index} $(\hat{\rho}=0.15)$.}
    \label{fig:Index-Trade-Simulated}
  \end{subfigure}
  \begin{subfigure}[b]{0.5\textwidth}
    \includegraphics[width=\textwidth]{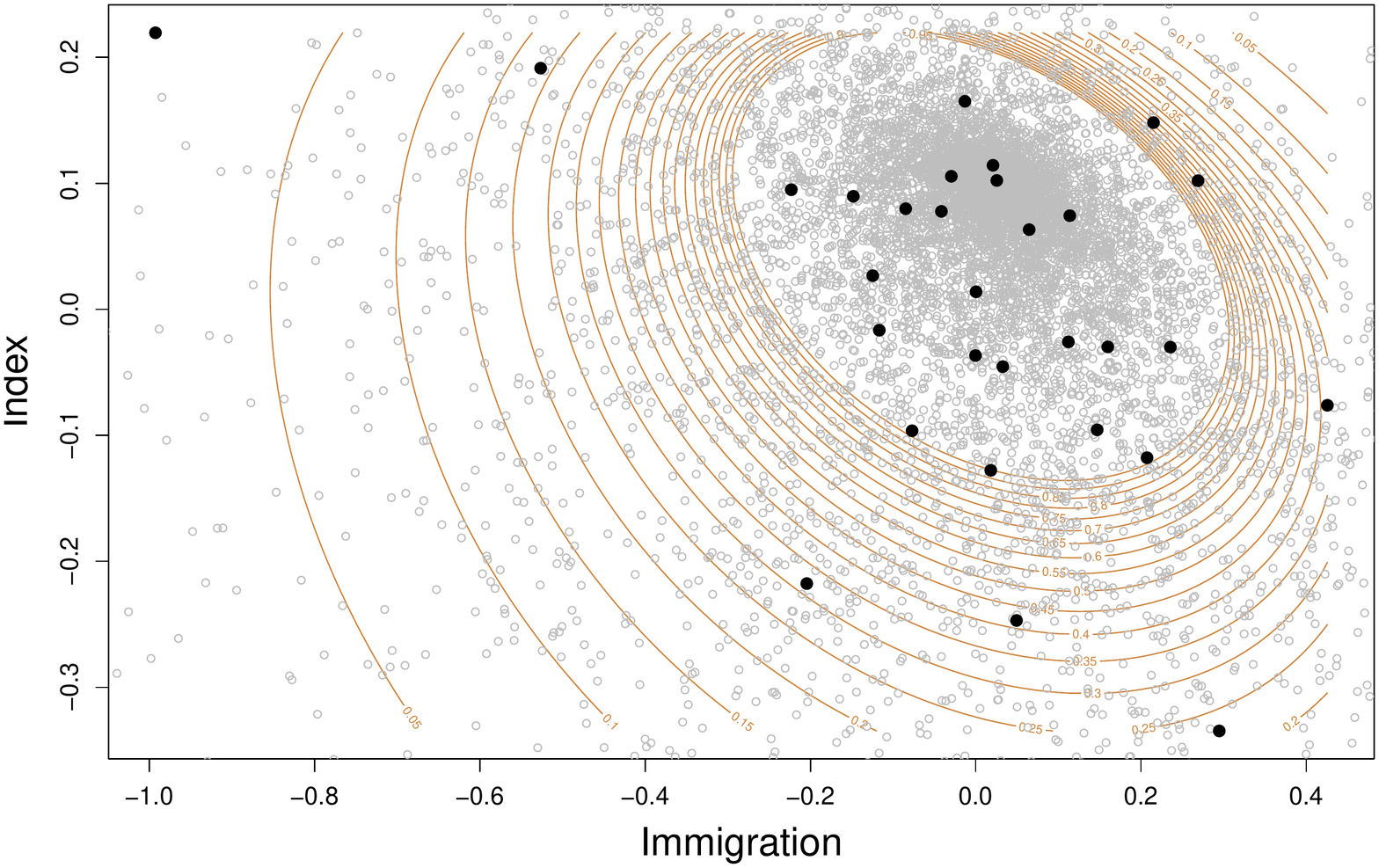}
    \caption{\texttt{Immig} vs.~\texttt{Index} $(\hat{\rho}=-0.44)$.}
    \label{fig:Index-Immig-Simulated}
  \end{subfigure}
\caption{Contour plots of fitted bivariate GH models to the joint
  log-returns of \texttt{Trade} and \texttt{Index} (left panel), and
  \texttt{Immig} and \texttt{Index} (right panel). The observed data
  and the 10,000 simulated values from the fitted model are displayed
  as black and grey dots, respectively.}
\end{figure}

Starting from these three plausibly iid
series, we proceeded by fitting bivariate GH models to their joint
marginal distributions: \texttt{Trade} vs.~\texttt{Index}, and
\texttt{Immig} vs.~\texttt{Index}. The models suggested by AIC and BIC
had values for the tail parameter of approximately $\lambda\approx
0.4$, which is somewhat close to a VG distribution, although the latter, as
well as NIG, provided substantially inferior fits. In order to
compute the systemic risk measures discussed below, 10,000 simulated values were
drawn from these models. Figures~\ref{fig:Index-Trade-Simulated}
 and \ref{fig:Index-Immig-Simulated} display the fitted contour plots
 from each model, overlaid with the 10,000 simulated values and the
 30 observed data points. As is noted in the figures, the empirical
 correlation coefficients based on the observed data, suggest a weak positive relationship between \texttt{Trade}
 and the \texttt{Index} ($\hat{\rho}=0.15$), and a moderate negative association between \texttt{Immig}
 and the \texttt{Index} ($\hat{\rho}=-0.44$).  

The assessment of how extremely negative events on the stress factors,
i.e., values on their left tail,
impact the \texttt{Index}, can be done with the various measures of
systemic risk currently in vogue, all of which target the left tail of
the conditional distribution of the random variable of interest, $Y$,
given that the stressor, $X$, is at or below some low quantile.
This notion of Conditional Value-at-Risk
(CoVaR), was originally introduced by \cite{covar0}.

Here, and due to the better properties pointed out by \cite{Mainik14}, we opt to take the variant of CoVaR developed by 
\cite{girardi2013systemic}. If $F_{Y|X}$ denotes the conditional
distribution of $Y$ given $X$, each with respective distribution
functions $F_{Y}$ and $F_{X}$, then we denote by $\xi_{q}$ the CoVaR
at level $q$, or $\text{CoVaR}_q$, which is defined to be 
\begin{equation}\label{eq:def-CoVaR}
\xi_q := \text{CoVaR}_q := F^{-1}_{Y|X\leq F^{-1}_{X}(q)}
\left( q \right) = \text{VaR}_{q}\left(Y|X\leq\text{VaR}_{q}(X)\right),
\end{equation}
where, using fairly standard notation from the risk modeling
literature, $\text{VaR}_{q}\left(X\right):=F^{-1}_{X}(q)$ denotes the
VaR of $X$ at
level $q$, which is simply the $q$-quantile of $X$. In line with this
concept, the analogous value for the closely associated Expected
Shortfall (ES), defined as the tail mean beyond VaR, is given by \citep{Mainik14}
\begin{equation}\label{eq:def-CoES}
\text{CoES}_q := \E\left(Y|Y\leq\xi_{q},X\leq\text{VaR}_{q}(X)\right).
\end{equation}
A variation on this proposed by \cite{ZariCOETL} is
\begin{equation}\label{eq:def-CoES}
\text{CoETL}_q := \E\left(Y|Y\leq\text{VaR}_{q}(Y),X\leq\text{VaR}_{q}(X)\right).
\end{equation}

The stress-testing results are presented in
Table~\ref{tab:risk-measures}. The general pattern seems to be that
for the same level, stress on \texttt{Trade} appears to have a marginally larger impact on
the \texttt{Index} than does stress on \texttt{Immig}; a finding which
is also consistent with the sign of the correlation coefficients noted
above. However, at the
highest stress level of 1\%, the results are mixed. 
\begin{table}[!htb]
\caption{
Left-tail systemic risk measures on the \texttt{Index} at different
levels, based on stressing the factors \texttt{Trade} and \texttt{Immig}.}
\label{tab:risk-measures}
\centering
\begin{tabular}{cc|ccc}
\toprule
Stress  & Stress & \multicolumn{3}{c}{Risk Measure on \texttt{Index}
                   (left tail)}
     \\ \cline{3-5}
Factor  & Level  & CoES & CoVaR & CoETL \\
\midrule
\multirow{3}{*}{\texttt{Trade}}
 & 10\% & -0.8403 & -0.5448  & -0.4980  \\
 & 5\%  & -1.2377 & -0.8414  & -0.6668 \\
 & 1\%  & -1.6367 & -1.2331  & -0.9728   \\
\midrule
\multirow{3}{*}{\texttt{Immig}} 
 & 10\% & -0.7027 & -0.4370 & -0.4748 \\
 & 5\%  & -0.9902 & -0.7288 & -0.6645   \\
 & 1\%  & -1.3788 & -1.2682 & -1.0084   \\
\bottomrule
\end{tabular}
\end{table}

\section{Conclusion and Discussion}
We proposed an annual well-being index (SWBI) constructed as the log-returns of 
an equally-weighted linear combination of several socioeconomic factors,
in order to dynamically measure the mood of US citizens. The data,
publically available from reliable government sources, spans the 30-year period 
from 1986 to 2016. Although the
SWBI exhibits no apparent serial dependence, we fitted an ARMA-GARCH
time series model in order to appropriately capture the marginal (or
cross-sectional) distribution which was consistent with a member of
the generalized hyperbolic family, as predicted by contemporary
best-practices financial theory \citep{massing2019best}. This procedure was also a
prerequisite step to generating valid option prices and performing
risk budgeting for the SWBI. The resulting values reveal the relationships among time
to maturity, strike price, and option price, enabling the construction
and valuation of insurance-type financial instruments.

To complete the rational finance-based valuation, we performed risk
budgeting for an equally weighted portfolio, and
 stress-tested the index by examining the effect of the exogenous but
 politically-sensitive variables, trade imbalance and amount of legal
 immigration, on the SWBI systemic risk. Among the component series
 of the SWBI, the VXO volatility measure of the stock market is
 the greatest contributor to tail risk. Among the external factors,
it appears that the level of
 trade imbalance tends to be associated with a larger impact on
 negative well-being than does immigration; a conclusion that mirrors the
 empirical finding that trade imbalance is positively correlated with SWBI, while
 immigration is negatively correlated. 

The main intent of the SWBI is for it to provide an early-warning mechanism
for downturns in the mood of US citizens. Coupled with the proper
valuation of SWBI financial instruments, it is hoped that this will
alert investors to potential future crises assess and aid them in
setting in place hedging strategies necessary to mitigate against possible losses.


\end{document}